# VARIABLE BLOCK CARRY SKIP LOGIC USING REVERSIBLE GATES

Md. Rafiqul Islam, Md. Saiful Islam, Muhammad Rezaul Karim, Abdullah Al Mahmud and Hafiz M. Hasan Babu
{rafik@udhaka.net}, {sohel_csdu@yahoo.com}, {r_karimcs@yahoo.com}, {aamrubel@yahoo.com}, {hafiz@hotmail.com}

Department of Computer Science and Engineering, University of Dhaka
Dhaka-1000, Bangladesh

**Abstract:** Reversible circuits have applications in digital signal processing, computer graphics, quantum computation and cryptography. In this paper, a generalized k*k reversible gate family is proposed and a 3*3 gate of the family is discussed. Inverter, AND, OR, NAND, NOR, and EXOR gates can be realized by this gate. Implementation of a full-adder circuit using two such 3*3 gates is given. This full-adder circuit contains only two reversible gates and produces no extra garbage outputs. The proposed full-adder circuit is efficient in terms of gate count, garbage outputs and quantum cost. A 4-bit carry skip adder is designed using this full-adder circuit and a variable block carry skip adder is discussed. Necessary equations required to evaluate these adder are presented.

## 1. INTRODUCTION

The input vector of reversible circuit can be uniquely recovered from the output vector, that is, for each input pattern there is a unique output pattern. Logic computations that are not reversible necessarily generate heat irrespective of their implementation technologies. According to [2], zero energy dissipation would be possible only if the network consists of reversible gates.

Synthesis of reversible logic circuits differs significantly from the synthesis of combinational (classical) logic circuits. Because in a reversible circuit the number of inputs must be equal to the number of outputs, every output can be used only once (i.e., no fan-out is permitted), and the resulting circuit must be acyclic.

Therefore, a good synthesis method must take into account the following rules:
1. use as many outputs of every gate as possible, and thus minimize garbage (unused) outputs.
2. do not create more constant inputs (required to make an irreversible specification to a reversible one) to gates than is absolutely necessary.
3. avoid leading output signals of gates to more than one input, because each fan-out of two requires adding one copying circuit.

The rest of the paper is organized as follows: section 2 presents the families of reversible gates and their quantum cost. Section 3 presents a generalized k*k reversible gate and discusses an instance of this family of gates. Section 4 first establishes the minimum number of constant inputs and garbages are required to design a full adder circuit, and then composition of a new full adder circuit is proposed. Section 5 presents the design of a carry skip adder using the proposed full adder circuit for which it is used as basic building block. Section 6 presents a variable block carry skip adder block. Experimental results are shown in section 7. Section 8 concludes the paper. References are listed in section 9.

## 2. FAMILIES OF REVERSIBLE GATES AND THEIR QUANTUM COST

There exist many universal reversible gates [1,3,7,10,11]. There exists only one 1*1 reversible gate called inverter ($A \rightarrow A'$). This gate is very important since it does not introduce garbage outputs. Some of the popular and important gates are 2*2 Feynman ((A, B)$\rightarrow$(P=A, Q=A$\oplus$B)), 3*3 Toffoli ((A, B, C) $\rightarrow$(P = A, Q=B, R=AB$\oplus$C), 3*3 Fredkin ((A, B, C) $\rightarrow$(P = A, Q=A'B$\oplus$AC, R=A'C$\oplus$AB)) and Peres [1] ((A, B, C) $\rightarrow$ (P = A, Q = A$\oplus$B, R=AB$\oplus$C)) gate.

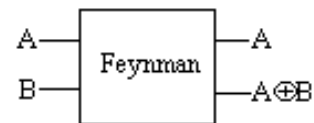

**Figure 1. 2*2 Feynman Gate**

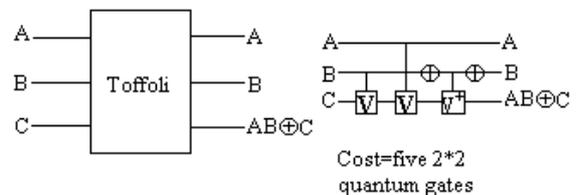

**Figure 2. 3*3 Toffoli Gate**

The detailed cost of a reversible gate depends on any particular realization technology of quantum logic. According to [9], it is assumed that the cost of every 2*2 is the same. A 1*1 cost nothing, since it can be always included to arbitrary 2*2 gate that preceded or follows it. Thus, every permutation quantum gate will be build from 1*1 and 2*2 quantum primitives and its cost calculated as a total sum of 2*2 gates.

Using the well known realization of Toffoli gate with truly quantum 2*2 primitives, according to [9], the cost of Toffoli gate is five 2*2 gates, or simply, 5 as shown in figure 2. The cost of Fredkin gate is exactly the same as the cost of Toffoli gate [5], which is shown in figure 3. Peres gate can be realized with cost 4 [9]. This is the cheapest quantum realization of a complete (universal) permutation gate.

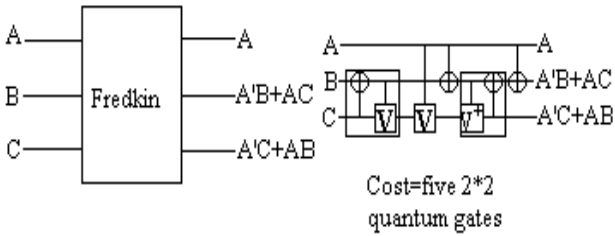

Figure 3. 3*3 Fredkin Gate

## 3. A GENERALIZED K*K REVERSIBLE GATE FAMILY

A generalized k*k reversible gate family is proposed in Figure 4(a), $f_{k-2}(A_1, A_2, ..., A_{k-2})$ is an arbitrary function of $A_1, A_2, ..., A_{k-2}$ and $f_{k-1}(A_1, A_2, ..., A_{k-1})$ is the function of $A_1, A_2, ..., A_{k-1}$. The gate is a (k-2) through gate. Mathematical properties of the gate family and systematic method for reversible logic synthesis using this family of gates are now being studied.

With k=2, this family of gate performs the same function as the **Feynman** gate. A 3*3 gate of the family is shown in Figure 4(b). The equation of this gate was known to **Peres** [1]. **The quantum cost of this circuit is 4**. The operation of this gate is shown in figure 5.

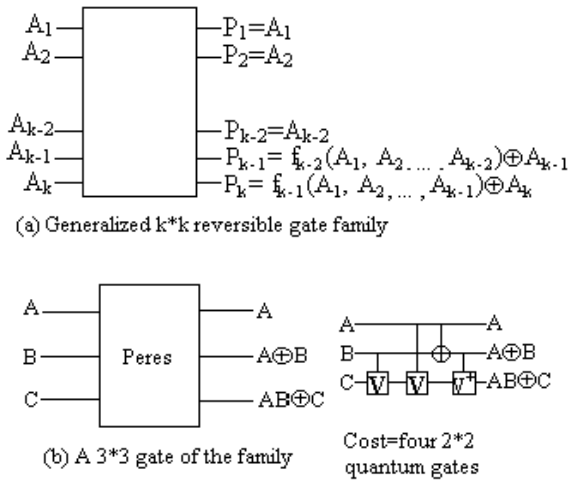

Figure 4. A generalized k*k reversible gate family and a 3*3 gate of the family

## 4. COMPOSITION OF FULL ADDER CIRCUIT

***Theorem:*** *A full-adder can be realized with at least two garbage output and one constant input.*

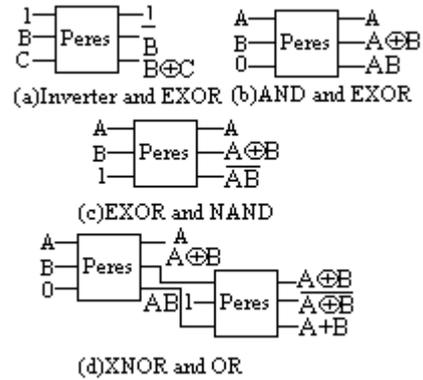

Figure 5. Operation of Peres gate

**Proof:**
The full-adder output S ($A \oplus B \oplus C_{in}$), Cout (($A \oplus B$)Cin $\oplus$ AB) equations produce the same output (1,0) for the three distinct input combinations (0,0,1), (0,1,0), and (1,0,0). Therefore, to separate all repeated values of outputs S and Cout we need at least two garbage outputs. Thus, a total of outputs is 2+2 = 4. Since in reversible circuits number of inputs must be equal to number of outputs and there are three inputs (A, B, and Cin), at least one constant input is necessary.

A full adder implementation using two 3*3 Toffoli gates and two 2*2 Feynman gate is presented in [8]. The circuit requires four reversible gates, produces two garbage outputs and the quantum cost is of 10.

Another full adder implementation using four 3*3 Fredkin gates is presented in [6] The circuit requires four reversible gates, produces two garbage outputs and the quantum cost is of 20.

In this paper, we present a new full adder composition. It consists only of two Peres gate and the quantum cost is of 8, which is minimum than all of the existing compositions. This we will call Peres full-adder which shown in figure 6.

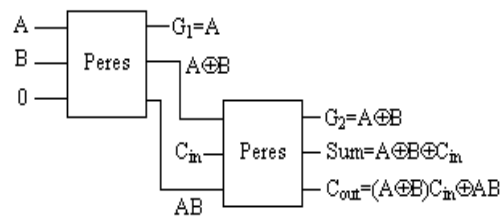

Figure 6. Peres Full Adder

## 5. CARRY SKIP-ADDER

The carry skip adder reduces the delay due to the carry computation. Consider the full-adder's operation. If either input is a logical 1, the cell will propagate the carry input to the carry output. Therefore, the $i^{th}$ full-adder carry input, $C_i$, will propagate the carry input to its carry output, $C_{i+1}$, when $P_i = A_i \oplus B_i$. Multiple full-adders, called a block, can generate a "block" propagate signal to detour the incoming carry around to the block's carry output signal. Figure 7 shows a 4-bit carry skip adder block. Each block is a small ripple carry adder producing the block's sum and carry bits.

However, each block quickly calculates whether the block's carry input is propagated to its carry output.

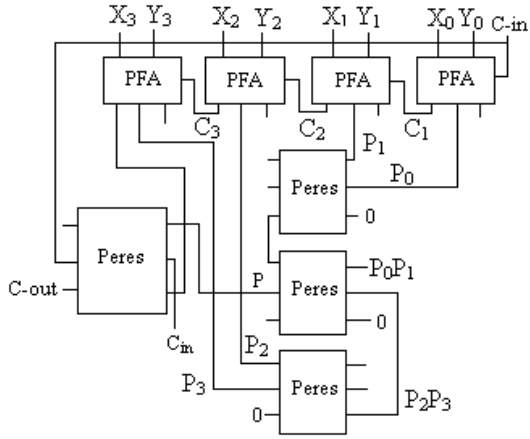

**Figure 7. Four Bit Carry Skip Adder**

A B bit full adder requires 2B Peres gate using the circuit in figure 6. A B input AND gate requires B-1 Peres gates. Therefore, a B bit carry skip adder requires 3B Peres gates.

Consider the B bit carry skip adder block in figure 7 generating a block carry out $C_{out}$ generates via carry ripple through the full adders. The least significant full adder requires a path delay of 2 Peres gates to generate $C_1$ from the addends. Then, the carry "ripples" through the subsequent full adders with a path delay of 1 Peres gate per bit. Finally, the Peres gate in the left of figure 7 generates Cout. Therefore, the delay to generate block carry out $C_{out}$ (via ripple) with a B bit carry skip adder is

$$d_{ripple}(B) = B + 1 \quad (1)$$

The full adder in figure 6 generate sum bit $S_i$, carry bit $C_i$, and propagate signal $P_i$ ($G_2$) passing through 2, 2, and 2 reversible gate. Therefore, the delay to generate $S_i$ is 2. The delay to generate $P_i$ is 2. And the delay to generate $C_i$ is 2 reversible gates. Then, all propagate signals for the carry skip adder block are combined with a B bit AND gate with delay $\lceil \log_2 N \rceil$. Finally, the Peres gate in the left of figure 7 generates Cout.

$$d_{skip} = \lceil \log_2 B \rceil + 3 \quad (2)$$

The total worst-case delay $T_{fixed}$ of an N bit carry skip adder with fixed block size B is the sum of the ripple carry delay through the first carry skip adder block, skip delays through the intermediate blocks and the ripple carry delay through the last block, or

$$T_{fixed} = (B+1) + \left(\frac{N}{B} - 2\right)(\lceil \log_2 B \rceil + 3) + (B+1) \quad (3)$$

Assuming $\lceil \log_2 B \rceil \approx B/2$ we get

$$T_{fixed} = B + \frac{N}{2} + \frac{3N}{B} - 4 \quad (4)$$

Minimizing $T_{fixed}$ with respect to block size B gives

$$1 - \frac{3N}{B^2} = 0 \text{ or } B_{opt} = 1.73\sqrt{N} \quad (5)$$

Substituting (5) into (4) gives the shortest delay for a fixed block size carry skip adder

$$T_{fixed} = \frac{N}{2} + 3.47\sqrt{N} - 4 \quad (6)$$

## 6. VARIABLE BLOCK CARRY SKIP ADDER

Varying the size of the carry skip adder blocks can reduce the total worst-case delay. Since carries generated or absorbed in the adder circuits have shorter data paths. Without loss of generality, the first and last carry skip blocks are b bits wide, and the carry skip adder is divided into t blocks, where t is even. Assuming the carry skip adder block sizes are

$$b \quad b+1 \quad \ldots \quad b+\frac{t}{2}-1 \quad b+\frac{t}{2}-1 \quad \ldots \quad b+1 \quad b \quad (7)$$

Summing the number of bits in the blocks, equating to N, and rearranging gives

$$b = \frac{N}{2} - \frac{t}{4} + \frac{1}{2} \quad (8)$$

The total worst case delay $T_{variable}$ of an N bit carry skip adder with the variable block sizes is the sum of the ripple carry delay through the first carry skip adder block, the skip delays through the intermediate blocks, and the ripple carry delay through the last block. Assuming the variable block sizes in (3), the total delay is

$$T_{variable} = 2(b+1) + 2\sum_{k=b+1}^{b+\frac{t}{2}-1} (\lceil \log_2 k \rceil + 3) \quad (9)$$

Assuming $\lceil \log_2 k \rceil \approx k/2$ and rearranging (9) becomes

$$T_{variable} = 4b - 4 + 3t + \frac{\left(b+\frac{t}{2}-1\right)\left(b+\frac{t}{2}\right) - b(b+1)}{2}$$

$$= \frac{t^2}{8} + \frac{11t}{4} + \frac{bt}{2} + 3b - 4 \quad (10)$$

Inserting (8) into (10), and collecting terms gives

$$T_{variable} = \frac{9t}{4} - \frac{5}{2} + \frac{3N}{t} + \frac{N}{2} \quad (11)$$

The optimal number of blocks is found with

$$\frac{\delta T_{variable}}{\delta t} = \frac{9}{4} - \frac{3N}{t^2} = 0 \text{ or } t_{opt} = \frac{2}{3}\sqrt{3}\sqrt{N} \quad (12)$$

Therefore, the optimal variable block size carry skip adder has delay

$$T_{variable} = \frac{N}{2} + \sqrt{3}\sqrt{N} - \frac{5}{2} \quad (13)$$

## 7. RESULTS

We compare our proposed full adder circuits with existing designs and result is shown in Table 1 and Table 2. In the previous paper Quantum costs of those circuits are not considered. We calculate the Quantum cost of those adder circuits and compare them with our proposed design.

**Table 1: Comparison Table1**

| Full-adder Composition | No. Of gates used | No. Of Garbage Output | No. Of Constant input | Quantum Cost |
|---|---|---|---|---|
| Proposed Peres | 2 | 2 | 1 | 8 |
| Toffoli, Khan and Feynman [4] | 3 | 2 | 1 | - |
| Toffoli and Feynman [8] | 4 | 2 | 1 | 10 |
| Khan and Feynman gate [7] | 3 | 3 | 2 | - |
| Fredkin [6] | 4 | 3 | 2 | 20 |

The analytical performance of the carry skip adder in [6] and our carry skip adder (Figure 7) is given in table 3. It is evident from Table 3 that our design performs better. For smaller block size our carry skip adder performs best (approximately double) and practically smaller block size is required. We choose binary exponential values for the block size, which is natural for block size.

**Tabe1. Comparison Table 2**

| Full-adder Composition | Unit Clock Cycle | Gate Calculations | | |
|---|---|---|---|---|
| | | Two input EXOR | Two input AND | NOT |
| Proposed Peres | 2 | 4 | 2 | 0 |
| Toffoli, Khan and Feynman [4] | 3 | 4 | 3 | 0 |
| Toffoli and Feynman [8] | 4 | 4 | 2 | 0 |
| Khan and Feynman gate [7] | 3 | 5 | 4 | 6 |
| Fredkin [6] | 4 | 8 | 16 | 4 |

## 8. CONCLUSION

The main goal of this paper is finding a good architecture for adder circuits using reversible logic based on minimizing gate count, garbage outputs, constant inputs, and quantum cost. Technology independent analysis of these adder circuits is given since quantum or optical logic implementations are not available.

**Table 3: Showing $T_{fixed}$ for different Implementations**

| No. Of Bits | $T_{fixed}$ for[6] | $T_{fixed}$ for Peres |
|---|---|---|
| 4 | 13.49 | 4.93 |
| 8 | 21.9 | 9.80 |
| 16 | 34.98 | 17.86 |
| 32 | 55.82 | 31.60 |
| 64 | 89.97 | 55.71 |
| 128 | 147.64 | 99.19 |
| 256 | 247.94 | 179.43 |
| 512 | 427.27 | 330.38 |
| 1024 | 755.87 | 618.85 |
| 2048 | 1370.54 | 1176.77 |
| 4096 | 2539.74 | 2265.70 |